\documentclass[aps,prl,a4paper,reprint,showpacs,amsmath,amssymb,superscriptaddress,floatfix]{revtex4-1}

\usepackage{amssymb}
\usepackage{latexsym}
\usepackage[dvips]{graphicx}
\usepackage{epsfig}

\usepackage{dcolumn}
\usepackage{amsmath}
\usepackage{amsfonts}
\usepackage{bm}
\usepackage{color}

\newcommand{\be}{\begin{equation}}
\newcommand{\ee}{\end{equation}}
\newcommand{\bea}{\begin{eqnarray}}
\newcommand{\eea}{\end{eqnarray}}

\newcommand{\p}{\partial}

\newcommand{\la}{\langle}
\newcommand{\ra}{\rangle}

\newcommand{\re}{\mbox{e}}

\renewcommand{\vec}[1]{{\bm #1}}

\begin{document}

\title{Chiral Lattice Supersolid on Edges of Quantum Spin Hall Samples}

\author{Oleg M. Yevtushenko}
\affiliation{Ludwig Maximilians University, Arnold Sommerfeld Center and Center for Nano-Science,
                 Munich, DE-80333, Germany}

\author{A. M. Tsvelik}
\affiliation{Condensed Matter Physics and Materials Science Division, Brookhaven National Laboratory, Upton, NY 11973-5000, USA}

\date{\today }

\begin{abstract}
We show that edges of Quantum Spin Hall topological insulators represent a natural platform for
realization of exotic supersolid phase. On one hand, fermionic edge modes are helical due to the
nontrivial topology of the bulk. On the other hand, a disorder at the edge or magnetic adatoms may produce
a dense array of localized spins interacting with the helical electrons. The spin subsystem is
magnetically frustrated since the indirect exchange favors formation of helical spin
order and  the direct one favors (anti)ferromagnetic ordering of the spins. At a moderately strong
direct exchange, the competition between these spin interactions results in the spontaneous breaking
of parity and in the Ising type order of the $z$-components at zero temperature.  If the total spin is
conserved the spin order does not pin a collective massless helical mode which supports the ideal
transport. In this case, the phase transition converts the helical spin order to the order of a chiral lattice
supersolid. This represents a radically new possibility for experimental studies of the elusive supersolidity.
\end{abstract}

\pacs{
   71.10.Pm,   
   73.43.-f       
   75.30.Hx,    
   67.80.K-,     
   67.80.kb      
}

\maketitle

Supersolid is an exotic phase where, very counterintuitively, crystal order  and an ideal transport coexist
in one and the same physical system \cite{boninsegni_colloquium_2012}.
Dating back to the 50-ties, the first discussions of supersolidity resulted in arguments against its existence
\cite{Penrose_Onsager_1956}. It was realized later that the quantum bosonic statistics could provide
all necessary conditions for formation of the supersolids and, starting from the 60-ties, the studies
were concentrated on  interacting bosons, in particular, on $^4$He
\cite{yang_1962,AndreevLifshitz,chester_1970,leggett_1970}.
It can crystallize at a high pressure and is expected to combine broken translational invariance with superfluidity.
In spite of  large interest and intense experimental efforts \cite{kim_Nat_2004,kim_Sci_2004}, the supersolid
phase has not been convincingly realized in helium. This failure calls for a search for alternative physical platforms
for supersolidity. Recent experiments aim at realizing supersolid in a system of cold atoms \cite{leonard_2017,li_2017}.
Another well-known alternative is provided by a possibility to have magnetic supersolid after mapping  the bosonic
theory onto a magnetic (or a quantum gas) lattice model \cite{matsuda_1970,liu_1973}, where both the spin rotation
symmetry and the lattice symmetry can be broken simultaneously \cite{murthy_1997,momoi_2000,tsurkan_2017}.
The  longitudinal- and the transverse components of the antiferromagnetic order of the magnetic lattice model (or
the diagonal- and the off-diagonal long-range order of the quantum gas lattice model) correspond respectively to
the crystalline order and to superfluidity of the bosons. The transition to the supersolid phase on the lattice can be
related to the Dicke- and to the Ising type transitions, cf. Refs.\cite{baumann_2010,romhanyi_2011}.

{\it In this Letter}, we suggest a novel platform for the lattice supersolid phase. It is provided by the
recently discovered time reversal invariant topological insulators \cite{HasanKane,QiZhang,TI-Shen} which have
become famous due to  their virtually ideal edge transport. We will concentrate on
two-dimensional topological insulators -- Quantum Spin Hall samples (QSH) -- where transport is carried by the so-called
one-dimensional (1D) helical edge modes. These modes  possess  lock-in relation between electron spin and momentum so that
helical electrons propagating in opposite directions have opposite spins \cite{WuBernevigZhang,XuMoore}. This locking
protects transport against disorder \cite{Molenkamp-2007,EdgeTransport-Exp1,EdgeTransport-Exp0}:
An elastic backscattering of the helical electron must be accompanied by a spin-flip and, therefore,
it can be provided only by magnetic impurities \cite{MaciejkoOregZhang}. However, a single Kondo impurity
is unable to change the ideal dc conductance \cite{FurusakiMatveev} if the total spin is conserved.
Under some conditions, e.g. a random anisotropy of the Kondo coupling, the ballistic conductance may be
suppressed if the helical electrons are coupled to a dense Kondo array \cite{CheiGlaz,AAY,Yevt-Helical,vayrynen_2016}.
The latter can be present in realistic samples due to the edge disorder which easily localizes a fraction of
the bulk electrons close to the edge \cite{AAY} such that the localized electrons become spin-1/2 local moments.
Alternatively, the disordered or regular Kondo array can be generated by magnetic adatoms located close to the edge,
cf. Ref.\cite{checkelsky_2012}.

\begin{center}
\begin{figure}[t]
   \includegraphics[width=0.48 \textwidth]{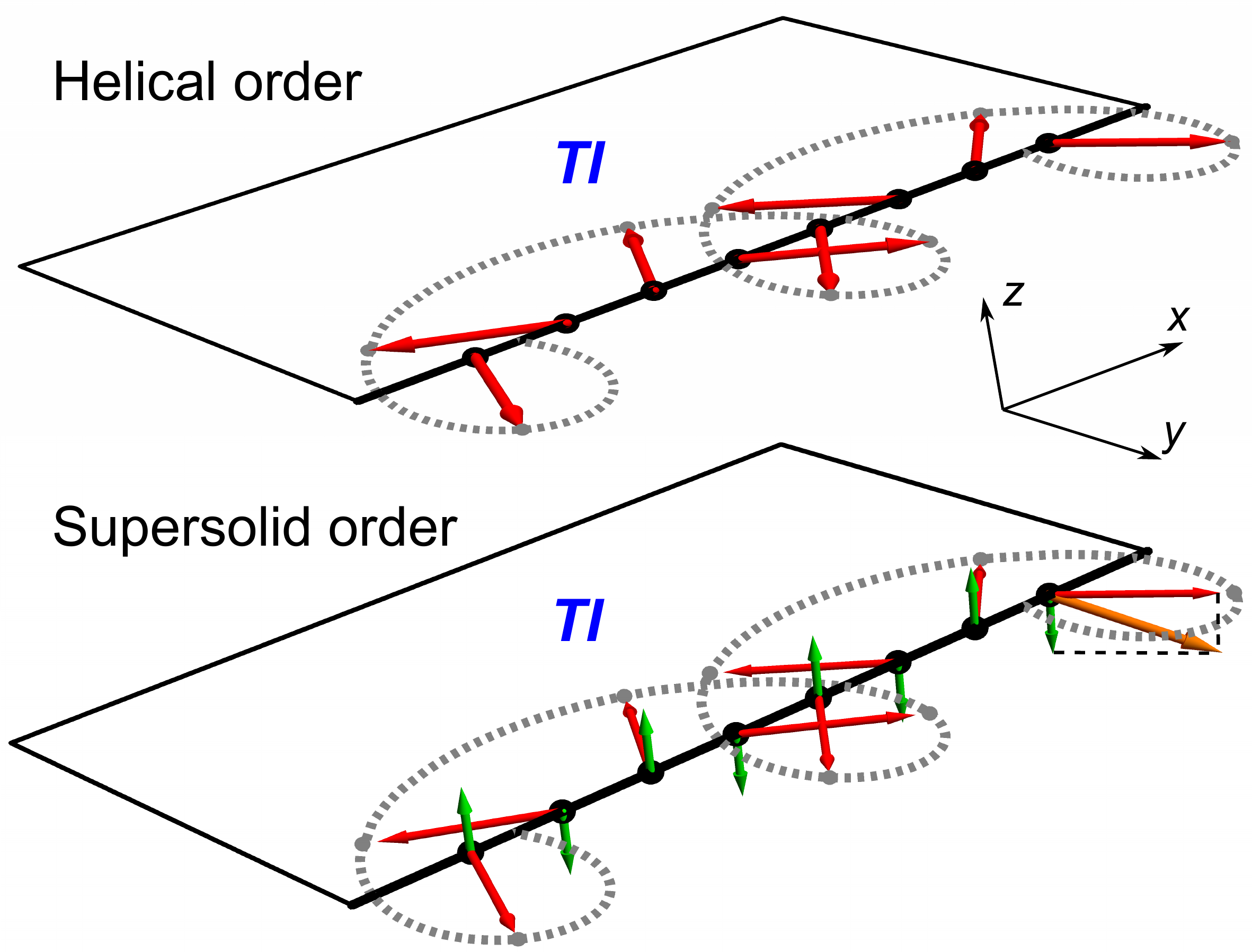}
   \caption{
   \label{SpinOrder}
        (color on-line) Illustration of the helical- (upper panel) and of the
        supersolid- (lower panel) spin order.
        Red- and green- arrows show in-plane- and z-components of the spins,
        respectively. The orange arrow is an example of the total spin orientation
        in the supersolid phase. Dotted line stands for the helix and black circles
        are the lattice sites. We have assumed that the isotropy of the $ xy $-plane
        is inherited from the bulk of QSH.
           }
\end{figure}
\end{center}

While transport of the helical 1D fermions coupled to a dense Kondo array has been intensively studied, magnetic
properties of these systems have attracted less attention. It is known that helical spin ordering, similar to that caused
by dynamical instabilities \cite{sun_helical_2015}, can result from the indirect Ruderman--Kittel--Kasuya--Yosida
(RKKY) spin interaction mediated by the helical electrons \cite{AAY,Yevt-Helical,VCSO}. This
holds true also when the electrons interact with each other and can be described as the Helical Luttinger Liquid (HLL).
However, to the best of our knowledge, the direct Heisenberg exchange interaction between the Kondo impurities has
never been taken into account though one may expect it to appear at relatively high spin densities. {\it We will show that},
if the Heisenberg coupling, $ J_H $, is sufficiently strong, the helical magnetic order on the QSH edge is converted
to another exotic magnetic state which we will call {\it Chiral Lattice Supersolid}, see Fig.\ref{SpinOrder}.


Our prediction is prompted by the recent theory for Kondo-Heisenberg models which states that a competition
of RKKY with the Heisenberg exchange may lead to the Ising-type phase transition \cite{KH-CSL}. Time-reversal
and parity symmetries are spontaneously broken in the ordered phase and, if the system is SU(2) symmetric,
spins form the isotropic scalar chiral spin order. It is characterized by an exotic local order parameter in
the form of a mixed product of three neighboring spins \cite{Zee89,Baskaran89}.

The lattice which we consider is very unusual:
SU(2) symmetry is broken at the QSH edges by helicity of the electrons. Therefore, the RKKY-Heisenberg competition
leads to the formation of a different exotic state. It combines (i) the helical transverse- and the (anti)ferromagnetic
spin orders; and (ii) the helical transport which is supported by collective modes of the helical electrons coupled to the
transverse spin fluctuations.
These modes are slow due to the strong electron-spin coupling. Even more importantly, they
are gapless, i.e. transport is ideal, provided the total spin is conserved.
One can say that the electrons play an auxiliary r{\'o}le for formation of the spin orders
but their helicity is crucially important for the ideal transport.

Both types of the spin order appear only at $T=0$. At finite temperature, the corresponding correlation lengths are finite
but diverge at $T \to 0$. Hence there is a region of temperatures $ T < \Delta $ where the correlation lengths
are large in comparison with the scale $ O(1/\Delta) $; $ \Delta $ is the characteristic energy scale below which the coupling
between the electrons and the localized moments becomes strong,  see Eq.(\ref{Delta}) below.
In this temperature range, the proximity to the ordered state is strongly felt and the spin order
is present. Thus, we come across all properties of supersolidity described above; the name ``Chiral Lattice Supersolid''
reflects their unique combination peculiar to the QSH edges. These are our {\it main results}. We note in passing that,
if $ T > \Delta $, the system is in a regime of a weak coupling between spins and electrons.

Now, we will introduce the model and explain the key steps of our approach and
the most important equations \cite{Comm-SkipAlgebra}. The Hamiltonian of  HLL
coupled to an array of  interacting localized spins is:
$ \, \hat{H} = \hat{H}_0 +  \hat{H}_{\rm int} + \hat{H}_H + \hat{H}_K $, where
the first two terms describe the free fermions and the interaction between
them, respectively:
\begin{eqnarray}
\label{H0}
       \hat{H}_0 & = & - i v_F \int {\rm d} x \
                                       \sum_{\eta = \pm} \eta \, \psi^{\dagger}_\eta(x) \partial_x \psi_\eta (x) , \\
\label{Hint}
       \hat{H}_{\rm int} & = & \frac{g}{2 \nu} \int {\rm d} x \, \left( \rho_+ + \rho_- \right)^2 \!\! ,
       \quad
       \rho_\pm \equiv \psi^{\dagger}_{\pm} \psi_{\pm} .
\end{eqnarray}
Here $ \, \psi_+ $ \, ($\psi_-$) describes spin-up right moving (spin-down left moving) in
the $x$-direction helical fermions $ \, \psi_{R,\uparrow} \, $ ($\psi_{L,\downarrow}$);
$ \, v_F $ is the Fermi velocity, $ \, \nu \, $ is the density of states in the HLL and
$ g $ is the dimensionless interaction strength which governs the Luttinger parameter $ K = 1/\sqrt{1+g} $.
We remind the readers that the electrostatic repulsion corresponds to  $ 0 < K < 1 $
and $ K = 1 $ denotes non-interacting electrons.

Without loss of generality, we consider the isotropic short range exchange interaction between
neighboring spins described by the Hamiltonian:
\be
\label{Hh}
 \hat H_H = J_H  \sum_{m} \vec{S}(x_{m + 1}) \vec{S}(x_m), \ x_m = \xi m;
\ee
with $ S_{x,y,z} $ being s-spin operators on the lattice sites $ x_m $. The sum runs over
sites of the spin array though, for the sake of simplicity, we will not distinguish constants
of the crystalline- and of the spin lattices, $ \xi $. Following the analogy with the magnetic
lattice supersolid, we choose the antiferromagnetic exchange, $ J_H > 0 $.

The most relevant coupling between the spins and the helical electrons is described by the
back-scattering Hamiltonian:
\begin{equation}
\label{Hb}
      \hat{H}_{K} = \int \!\! {\rm d} x \, \rho_s J_K
       \left[ S^+ e^{2 i k_F x} \psi^{\dagger}_{-}\psi_{+} + h.c. \right] \!\! ;
\end{equation}
where $ k_F \, $ is the Fermi momentum; $ J_K $ is $xy$-isotropic coupling constant; $ S^\pm
\equiv S_x \pm i S_y $. The dimensionless impurity density $ \, \rho_s \, $ has been used
to convert the sum over the lattice sites to the space integral.
We omit the forward-scattering term $ \, \sim J_z S_z \, $ since a unitary transformation
of the Hamiltonian allows one to map the theory with the parameters $ \, \{ K, J_z \ne 0 \} \, $
to the equivalent theory
with the effective Luttinger parameter $ \, \tilde{K} = K ( 1 - \xi J_z \nu / 2 K )^2 $ and $ \, \tilde{J}_z = 0 $
\cite{EmKivRes,MaciejkoLattice}. Thus, $ \, H_{\rm int} \, $ is able to take into account both the direct
electron-electron interaction and the interaction mediated by the $z$-coupling to the Kondo impurities.
The coupling constants are assumed to be small, $ s J_{H,K} \ll u/\xi, D $. Here $ D $ is
the UV energy cutoff which is of the order of the bulk gap in the QSH sample and
$ u $ denotes the excitation velocity renormalized by the electron interaction.

The model Eqs.(\ref{H0},\ref{Hint},\ref{Hb}) with $J_H =0$ was studied in Ref.\cite{Yevt-Helical}.
We will now briefly recapitulate key points of that paper and will generalize it for for finite $ J_H $.
Our goal is to to derive the effective low energy theory.
This can be conveniently done after  parameterizing the spins by unit vectors:
\bea
   S^{\pm}(x_m) & = & s \sqrt{1 -n^2_z(x_m)} \, \re^{\mp2 i k_F x_m \pm i \alpha(x_m)},  \cr
   S^z(x_m) & = & (-1)^m s \, n_z(x_m)  \, .
\label{spins}
\eea
Here, we have singled out slow spin variables $ \alpha, n_z $.
Next,  we change from the Hamiltonian to the action. Note that the parametrization Eq.(\ref{spins}) requires
the usual Wess-Zumino term in the Largangian \cite{ATsBook}, $ {\cal L}_{\rm WZ} = - i s \rho_s n_z \p_\tau \alpha$;
where $ \tau $ is the imaginary time. Using Eq.(\ref{spins}) and performing the gauge transformation of the fermionic
fields:  $ \psi_\eta e^{- i \eta \alpha/2} \to \psi_\eta $, we reduce the noninteracting fermionic part of the
Hamiltonian,  Eqs.(\ref{H0},\ref{Hb}), to the following  Lagrangian density:
\bea
   \label{Lagr0}
   {\cal L}_0 & = & \!\! \sum_{\eta=\pm} \!\!
                           \left[
                      \bar{\psi}_\eta \partial_\eta \psi_\eta + s \rho_s J_K \sqrt{1- n_z^2} \bar{\psi}_{-\eta} \psi_\eta
                           \right] \! + \! \frac{{\cal L}_{\rm LL}[\alpha,v_F]}{4}; \cr
  & & {\cal L}_{\rm LL}[\alpha, v_F] \equiv [ ( \p_\tau \alpha )^2 +   ( v_F \p_x \alpha )^2 ] / ( 2 \pi v_F ) \ .
\eea
Here $ \partial_\eta \equiv \p_\tau - i \eta v_F \p_x $ denotes the chiral derivative and $ {\cal L}_{\rm LL} $
is the hydrodynamic Lagrangian of the standard Luttinger liquid model.  $ {\cal L}_{\rm LL} $ has been generated by
the anomaly of the fermionic gauge transformation. Eq.(\ref{Lagr0}) contains the bare velocity $ v_F $ and no
Luttinger parameter $ \tilde{K} $ because we have not yet taken into account the electron interactions.

If we substitute a mean value $ {\cal M } \equiv \la \sqrt{1- n_z^2} \ra = {\rm const} $ for $ \sqrt{1- n_z^2 }$
the electrons acquire a constant gap in the spectrum, $ \Delta_0 = \bar{\Delta}_0 {\cal M} $ with $ \bar{\Delta}_0
\equiv  s \rho_s J_K $, which is opened by the back-scattering, Eq.(\ref{Hb}). By combining
the functional bosonization approach \cite{Yurkevich-2004} with self-consistent scaling arguments, one can show
that the main effect of the weak electron interaction, $ | \delta K | \ll 1 $ with $ \tilde{K} \equiv 1 - \delta K $,
is renomalization of the velocity, $ v_F \to u $, of the Luttinger liquid parameters,
$ {\cal L}_{\rm LL}[\alpha, v_F] \to {\cal L}_{\rm LL}[\alpha, u] / \tilde{K} $, and of the gap $ \Delta_0 \to \Delta \ll D$:
\be
\label{Delta}
  \frac{\Delta}{D} \simeq \left( \frac{\Delta_0}{D} \right)^{ \frac{1}{2-\tilde{K}} } \!\!\! \simeq
       {\cal M} \Bigl[ 1 - \delta K \log({\cal M}) \Bigr] \left( \frac{\bar{\Delta}_0}{D} \right)^{ \frac{1}{2-\tilde{K}} }  \!\!\! .
\ee
We will not consider the case $ {\cal M} \to 0 $ and, therefore, correction $ O[ \delta K \log({\cal M}) ] $ can be
safely neglected in Eq.(\ref{Delta}). To take into account stronger interactions, one could
employ the full (abelian) bosonization \cite{Bosonisation} and use the exact solution of the quantum sine-Gordon
model \cite{Zam1995,LukZam1997}; we will present this technically nontrivial extension elsewhere.

It is known that $ \alpha $ is gapless at $ J_H = 0 $ if the total spin is conserved \cite{AAY,Yevt-Helical}. We will
show that this statement holds true even at finite $ J_H $. Thus, Eq.(\ref{Lagr0}) describes
the connection between gapped- and gapless sectors which is mediated by (slow) fluctuations of $ n_z $.  In
other words, the energy scale $ \Delta $ establishes  a crossover from the weak- to strong coupling between the
electrons and the spins. In the strong coupling regime, they form a single Luttinger liquid where the low energy
charge excitations and the in-plane spin excitations are described by the same field $\alpha$ \cite{AAY,Yevt-Helical}.
Terminology used for the magnetic lattice supersolids suggests that $ {\cal M } $ will play in our approach the
role of the superfluid density.

Transition between the helical phase and supersolid can be identified after treating $ n_z $ and $ \alpha $ as
the slow variables and integrating out the gapped fermions. This yields the density of the  effective potential $ {\cal E}({\cal M}) $
per one unit cell. Restoring now finite $ J_H $, we find in the leading order in $ s J_K / D $:
\bea
\label{GS-En}
  {\cal E}({\cal M}) & \simeq & - \bigl( \xi \Delta^2 / 2 \pi u \bigr) \log(D/\Delta) + \\
                 & + & s^2 J_H {\cal M }^2 \Bigl(  1 + \cos[ 2 k_F \xi ]  \Bigr) + {\rm const};
  \nonumber
\eea
gradient terms will be discussed later. Minima of this effective potential determine
the ground state configuration of the magnetization field, $n_z$ .

If $ J_H $ is smaller than the critical value $ J_H^* $, the minimum is at $ {\cal M} = {\cal M}_h = 1 $
(i.e. $ \la n^z \ra =0$). The spins are in the $xy$-plane, see the upper panel of Fig.\ref{SpinOrder}.
When the Heisenberg exchange exceeds the critical value a nontrivial minimum appears at $ {\cal M}
= {\cal M}_s <1 $:
\be
  {\cal M}_s = \frac{D}{\bar{\Delta}}
                     \exp\left\{ \! - 4 \pi s^2 \frac{J_H u}{\xi \bar{\Delta}^2} \cos^2( k_F \xi_0 ) - \frac{1}{2} \right\} ;
  \label{Ms}
\ee
where $ \bar{\Delta} = \Delta / {\cal M} $ is the $ {\cal M} $-independent part of $ \Delta $.
Since $ {\cal M} \le 1 $, the origin of the nontrivial minimum requires $ J_H > J_H^* $; the
critical value is  defined by the equation
\be
\label{Jc}
     {\cal M}_s(J_H^*) = 1 \ \Rightarrow \ J_H^* \simeq
                        \frac{ \xi \bar{\Delta}^2  \log\left( D / \bar{\Delta} \right) }
                               { 4 \pi s^2 u \cos^2( k_F \xi_0 ) } .
\ee
We remind the readers that the Heisenberg coupling is assumed to be small. Therefore, the nontrivial
minimum can be realized only if $ s J_H^* \ll u/\xi, D $. This inequality implies that, in particular, the case
$ \cos^2( k_F \xi_0 ) \to 0 $ must be excluded from the consideration.

The solution Eq.(\ref{Ms}) corresponds to the staggered magnetization, see the lower panel
of Fig.\ref{SpinOrder}. Since $ \cal{E}({\cal M}) $ is invariant with respect to inverting the
spin components $ S_z $, $ n_z(m) \to -n_z(m) $ for all lattice sites,
the ground state is double-degenerate. This degeneracy is lifted
at $T=0$ by a spontaneous breaking of the corresponding $ Z_2 $ as in  1D
Ising model.

With a further increase of $ J_H $, the system approaches the isotropic Heisenberg magnet
where our approach requires significant modification. For the purposes of this paper, it is sufficient
to point out the existence of the lattice supersolid  phase at intermediate $ J_H $.

Fluctuations of $S^z$ are gapped for all values of $ J_H $ excluding  its critical value
$ J_H^* $. Therefore, the corresponding correlation functions are short ranged. On the other hand,
if the total spin is conserved, the effective action contains only gradients $ \p_\tau \alpha $ and
$ \p_x \alpha $, i.e., fluctuations of $\alpha$ {\it always are gapless}. The effective action for
$\alpha$, $ {\cal L}_\alpha $, can be derived by integrating out all massive modes: the fermions
and the $ n_z $ fluctuations \cite{AAY,Yevt-Helical}. This yields for the energies below
$ \Delta $:
\be
  {\cal L}_{\alpha} = {\cal L}_{\rm LL}[ \alpha, u_\alpha] / 4 K_\alpha.
\ee
One can show that  $ u_\alpha / K_\alpha \simeq u / \tilde{K} $ though the parameters of $ {\cal L}_\alpha $
are substantially renormalized by the electron-spin interactions such that $ K_\alpha \ll \tilde{K} $
and $ u_\alpha \ll u $. One can say that the massless excitations of our model are slow spinons
dressed by localized electrons. They govern the spin-spin correlations at $ T \ll \Delta $:
 \bea
&& \la\la S^+(\tau,x) S^-(0,0)\ra\ra \sim
          {\cal M}^2\re^{ -2 i k_F x} \la \re^{ i [ \alpha(\tau,x) - \alpha(0,0) ] } \ra = \cr
&& = {\cal M}^2 \re^{ -2 i k_Fx} \left[
                  \frac{ (\pi T \xi / u_\alpha )^2 }{ \sin^2 ( \pi T\tau ) + \sinh^2 ( \pi T x / u_\alpha ) }
                                                    \right]^{K_\alpha} \!\!\!\!\! .
\label{SS}
\eea
At $T=0$, the correlations in Eq.(\ref{SS}) decay as power law which is a signature of  a quasi long range
order of these components. The correlations are cut by the thermal length, $ L_T =  u_\alpha / T $, if the
temperature is finite.

\begin{figure}[t]
\begin{center}
   \includegraphics[width=0.48 \textwidth]{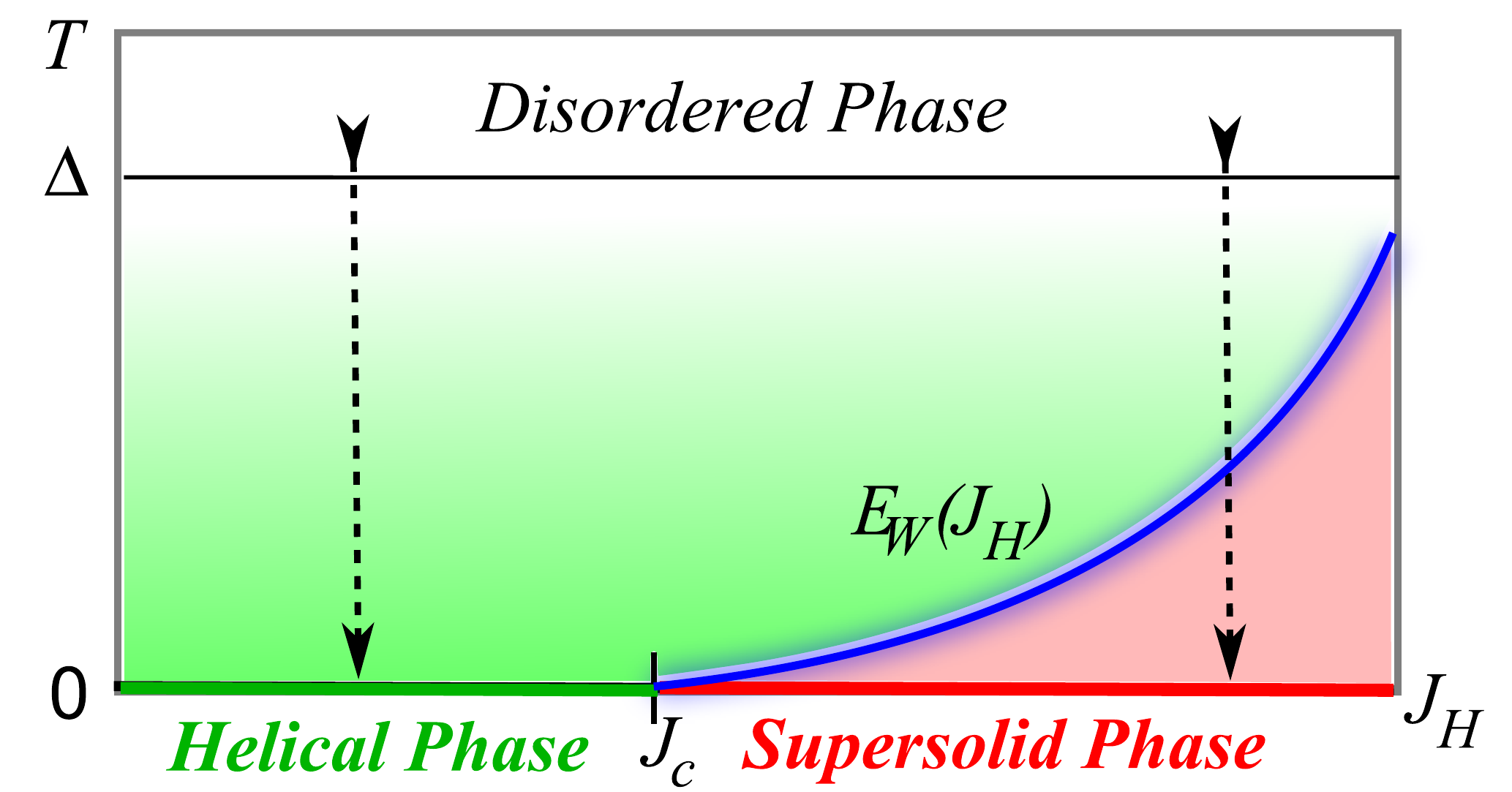}
\end{center}
\vspace{-0.25 cm}
   \caption{
        \label{PhaseDiagr}
        (color on-line) Phase diagram of the dense Kondo-Heisenberg array coupled to the interacting
        helical fermions at the edge of the Quantum Spin Hall topological insulator.
        Green- and red lines show phases with the helical- and supersolid order at $ T=0 $, respectively.
        Light-green- and light-red regions mark regimes where these orders are felt at finite $ T $. The
        supersolid order disappears at $ T \sim E_W $, see Eq.(\ref{Ewall}).
        The system becomes completely disordered at $ T \sim \Delta $, see Eq.(\ref{Delta}).
        Dashed lines exemplify protocols of measurements which could allow one to identify
        all phases, see the discussion in Conclusions.
           }
\end{figure}

{\it Helical phase}, $ J_H < J_H^* \mbox{ and } {\cal M} = {\cal M}_h = 1 $: The spin-spin correlation function
of $S^{\pm}$ components is given by Eq.(\ref{SS}) with fluctuations being centered at the wave vector $-2k_F$ (not
at $+2k_F$). This asymmetry is bound to the certain helicity of the fermions at the edge of QSH, see Eq.(\ref{Hb}):
helicity of the fermions governs orientation (right- or left handed) of the spin helix. The phase  has a nematic
(or vector chiral) order parameter reflecting the helical spin structure:
\be
\label{Oh}
  \overrightarrow{{\cal O}}_h = [ \vec{S}(x) \times \vec{S}(x+\xi) ], \
  [ \overrightarrow{{\cal O}}_h]_z \sim s^2\sin(2k_F \xi) .
\ee
The helical order is felt at $ \xi \ll u / \Delta \ll L < L_T $ where $  L $ is the system size. It becomes
suppressed at $ u / \Delta \ll L_T < L $ and is completely destroyed by the thermal fluctuation at
$ T \sim \Delta $, see Fig.\ref{PhaseDiagr}.

{\it Chiral Lattice Supersolid  phase}: $ J_H > J_H^* \mbox{ and } {\cal M} = {\cal M}_s < 1 $: in
addition to the helical order, a new order appears in the system via the Ising type transition, namely,
$ \la S_z \ra $ becomes staggered. Comparison with the theory of the magnetic lattice supersolid suggests
that this order is the counterpart of the broken translation symmetry. Since two spin orders coexist
with the gapless excitations, this phase is the lattice supersolid. This concludes the proof of our main result.

We emphasize that the new supersolid has some special features inherited from the helical phase.
The excitations are again centered at $-2k_F$ and not $2k_F$ [see Eq.(\ref{SS})] and,
therefore, are helical. The origin of this asymmetry is the same: nontrivial topology of the QSH bulk
which results in a certain helicity of the edge fermions. Moreover, the combination of the helical order
with the staggered magnetization trivially produces non-zero scalar chiral order parameter:
\be
  {\cal O}_c  =  \Big(\vec{S}(x-\xi), \overrightarrow{{\cal O}}_h  \Big).
\ee
To emphasize this complex nature of the new phase, we will refer to is as "Chiral Lattice Supersolid". We
note that the QSH samples are probably the unique platform where the 1D Chiral Lattice Supersolid
can be realized.

The finite temperature suppresses the staggered magnetization and, correspondingly, the supersolid order
via formation of domain walls. The energy of the single wall can be estimated
by the height of the potential barrier in the effective potential $ {\cal E}( {\cal M} )$:
\be
\label{Ew-General}
  J_H > J_c: \quad E_W \sim {\cal E}({\cal M}_h) - {\cal E}({\cal M}_c) .
\ee
For $ J_H $ close to $ J_c $, Eq.(\ref{Ew-General}) can be reduced to:
\be
\label{Ewall}
   E_W \sim
            \left[
               (J_H - J_c)/\bar{\Delta}
            \right]^2 \times ( \xi / u ).
\ee
The supersolid order can be felt if $ T < E_W $, see Fig.\ref{PhaseDiagr}, which ensures the
exponentially large correlation length of the field $ n_z $: $ L_{z} \propto \exp( E_W / T ) $.
The $Z_2$ symmetry is restored beyond the sale $ L_{z} $.

{\it To summarize}, we have demonstrated that, being coupled to a dense array of localized
quantum  spins, helical edge modes of a Quantum Spin Hall topological insulator can host
an exotic magnetic order at $T=0$. The system possesses a characteristic energy scale $\Delta$
related to the backscattering of the helical electrons from the local spins. This energy scale
signifies a crossover from  weak to  strong coupling. In the strong coupling regime the system
remains critical, but the spin fluctuations are absorbed into the electronic ones.

The temperature region  $ T < \Delta $ can be characterized by the proximity to the helical
spin order existing at $T=0$. Its underlying mechanism is based on the RKKY interaction of the
spins mediated by the helical electrons. More interestingly, a competition of the RKKY indirect
exchange with the direct Heisenberg one may lead at $ T=0, J_H > J_H^*$ [see Eq.(\ref{Jc})] to
the Ising type phase transition and to the appearance of of the additional order which is the
staggered magnetization. If the total spin is conserved these two spin orders coexist with gapless
excitation being able to support a virtually ideal transport. We have shown that there is one-to-one
correspondence between the new phase and the magnetic lattice supersolidity. Thus, the phase
which we have described is also a kind of the lattice supersolid which inherits peculiar features
of the helical magnetic phase. The latter has the nontrivial vector chiral order parameter,
Eq.(\ref{Oh}). That is why supersolid hosted by QSH samples can be called "Chiral Lattice Supersolid".

It is important that, if the total spin is conserved, a weak disorder in the spin lattice is not able
to suppress either the helical spin order or the ideal transport of the gapless excitations \cite{AAY}.
Clearly, the staggered magnetization can also appear in the weakly disordered Kondo-Heisenberg array
coupled to HLL. Thus, such a disorder can lead only to some quantitative changes and is unable to destroy
the Chiral Lattice Supersolid.

Our findings suggest that magnetically doped QSH edges provide a principally new
possibility for the realization and for the study of elusive supersolidity.
Coupling constants $ J_{K,H} $ can be controlled by varying the proximity of the 
magnetic adatoms to the helical edge and their density, respectively.
Experimental detection of the Chiral Lattice Supersolid can be based on spin-spin correlations,
i.e. spin susceptibilities, which have no pronounced peaks in the disordered phase. In the proximity
to the helical phase [see the left dashed line in Fig.\ref{PhaseDiagr} at $ T < \Delta$], correlation
functions of $ xy $-spin components acquire peaks at the wave vector $ Q_h = \pm 2 k_F $ with the sign
being defined by helicity of the electrons. The correlation function of $ z $-components is expected to be
structureless in the helical phase but must show new peaks at the Neel vector, $ Q_a = \pi / \xi $, in the
proximity to the supersolid phase [see the right dashed line in Fig.\ref{PhaseDiagr} at $ T < E_W $].
Thus, measuring the spin susceptibilities at different temperatures can fully characterize the system.

We have considered purely 1D system and, therefore, the spin order is only algebraic even in
the limit $ T \to 0 $. One promising generalization could include the study of the
Kondo-Heisenberg array coupled to the 2D edge of a 3D topological insulator. The influence of
fluctuations is weaker in 2D and, if the lattice supersolid can be realized in this setup, its
spin order is expected to become long-range.

\begin{acknowledgments}
{\bf Acknowledgments}:
O.M.Ye. acknowledges support from the DFG through the grant YE 157/2-1.
A.M.T. was supported by the U.S. Department of Energy (DOE), Division of Materials Science,
under Contract No. DE-AC02-98CH10886.
We are  grateful to Boris Altshuler for useful discussions.
\end{acknowledgments}

\bibliography{Bibliography,Supersolids,TI-Magn,F_Notes}

\begin{thebibliography}{45}
\bibitem [{VCS()}]{VCSO}%
  \BibitemOpen
  \href@noop {} {}\bibinfo {note} {We retain the term ``helical'' to reflect
  the relation between the spin order and the electron helicity. In the
  literature devoted to exotic magnetic phases, this order is often referred to
  as ``the vector chiral spin order''.}\BibitemShut {Stop}%
\bibitem [{Com()}]{Comm-SkipAlgebra}%
  \BibitemOpen
  \href@noop {} {}\bibinfo {note} {Though we have given a full quantum
  consideration to the derivation of the effective low energy theory there is
  no need to present details of the secondary importance; one can restore them
  straightforwardly with the help of
  Refs.\cite{Yevt-Helical,KH-CSL}.}\BibitemShut {Stop}%
\end{thebibliography}


\end{document}